\documentclass[intlimits,twoside,a4paper]{article}

\usepackage{amsmath,amssymb}
\usepackage{graphicx}
\usepackage[T2A]{fontenc}
\usepackage[cp1251]{inputenc}

\usepackage{cmpj3}

\issue{2019}{22}{2}{23801}
\doinumber{10.5488/CMP.22.23801}

\title[Linear modified Poisson-Boltzmann equation]
{Comments on the linear modified Poisson-Boltzmann equation in electrolyte solution theory}

\author[C.W. Outhwaite, L.B. Bhuiyan]{C.W. Outhwaite\refaddr{label1}, L.B. Bhuiyan\refaddr{label2}}

\addresses{
\addr{label1}Department of Applied Mathematics, University of Sheffield,
Sheffield S3 7RH, UK
\addr{label2}Laboratory of Theoretical Physics, Department of
Physics, University of Puerto Rico, \\
17 Avenida Universidad, STE 1701, San Juan, Puerto
Rico 00925-2537, USA
}

\date{Received March 26, 2019, in final form May 20, 2019}

\begin{document}

\maketitle

\begin{abstract}

Three analytic results are proposed for a linear form of the
modified Poisson-Boltzmann equation in the theory of bulk electrolytes.
Comparison is also made with the mean spherical approximation results.
The linear theories predict a transition of the mean electrostatic
potential from a Debye-H\"{u}ckel type damped exponential to a damped
oscillatory behaviour as the electrolyte concentration increases
beyond a critical value. The screening length decreases with increasing
concentration when the mean electrostatic potential is damped oscillatory.
A comparison is made with one set of recent experimental
screening results for aqueous NaCl electrolytes.


\pacs 82.45.Fk, 61.20.Qg, 82.45.Gj
\end{abstract}

\section{Introduction}

Little attention has been paid to the linear modified
Poisson-Boltzmann (MPB) equation in the electrolyte solution theory,
mainly due to the ready availability of the mean spherical approximation
(MSA) analytical results \cite{blum}. By contrast, the non-linear MPB
approach along  with the hypernetted chain (HNC) \cite{friedman}
theory have proved two of the more successful theories in predicting
the structure and  thermodynamic properties of the primitive model (PM)
(arbitrary sized  charged hard spheres moving in a continuum dielectric)
electrolyte solution  \cite{vlachy,bhuiyan,adriel}. It can be shown that
the MSA is the linearized  version of the HNC theory (see for example,
the references \cite{henderson1,henderson2}). However, unlike the MSA,
the linear MPB theory remains relatively less well explored.

In a broad sense, the usefulness of a linear theory is twofold.

\emph{Firstly}, from a theoretical point of view, linear theories
can and often provide valuable insights into the physics and chemistry of
a situation, which might otherwise remain obscure when analyzed through their
non-linear  counterparts only. A case in point is the similarity (or even
equivalence)  between the Debye-H\"{u}ckel (DH) parameter $\kappa $ \cite{debye}
and the analogous MSA parameter $\Gamma=[\sqrt{(1+2\kappa a)}-1]/(2a)$
($a$ is the ionic diameter) \cite{blum}. Although the HNC gives a more accurate
solution generally, it does not lead to a physically significant quantity such as
$\Gamma $. Similarly,  in our case, the linear MPB (LMPB) predicts the asymptotic
form of the non-linear  equation and gives  useful information regarding the
screening length. Being based on the mean  electrostatic potential approach,
the linear theory is a logical extension  of the Debye-H\"{u}ckel (DH) theory
\cite{debye}. In particular, a natural  transition is given of the mean electrostatic
potential behaviour in  going from a DH damped exponential to a damped oscillatory
as the concentration increases beyond some critical value.

\emph{Secondly}, from a practical perspective, linear
solutions are normally easier to obtain and are often analytic. The latter
feature makes a linear solution particularly convenient to use as an initial
guess in an iterative algorithm to obtain the corresponding non-linear solution.

    Screening has played an important role in analysing charged systems since
the electrical double layer work of Gouy and Chapman \cite{gouy,chapman}, and that
of Debye and H\"{u}ckel \cite{debye} in the electrolyte bulk. The phenomenon is
significant in shaping the structure and thermodynamics in these systems. A
recent theoretical work on the subject is due to Rotenberg et al. \cite{rotenberg}.
The experimental screening length decreases at the higher electrolyte
concentrations in contrast to that predicted by the DH theory. Smith et al.
\cite{smith}, Lee et al. \cite{lee} have recently investigated the  experimental
screening length of concentrated electrolytes. A scaling analysis \cite{lee}
suggests that the decay length increases linearly with the Bjerrum length at
 higher concentrations. When applied to the restricted primitive model (RPM),
the MSA and LMPB theories both predict a decrease in the screening length above
a critical $y_\text{c}$ ($y = \kappa a$).  Neither  theory can predict the scaling
result, a probable factor being the neglect of solvent effects.

	In this paper we will present the mean electrostatic potential results due to
three different versions of the LMPB, and compare them with the corresponding
results from the MSA, the MPB, and the DH theories. The comparative behaviour
of the screening properties predicted by the MPB and MSA theories will also be
examined.

\section{MPB theory}

Improvements to the classical PB theory rest upon a more accurate pair
distribution function $g_{ij}$ for two ions $i$ and $j$. In the mean electrostatic potential
approach, this implies an improvement in the mean electrostatic potential $\psi (1;2)$ at
the field point {\bf r}$_{2}$ for an ion $i$ at {\bf r}$_{1}$ through the solution of the
Poisson equation
\begin{equation}
\nabla ^{2}\psi _{i}(1;2)=-\frac{1}{\varepsilon _{0}\varepsilon_\text{r}}\sum _{s}e_{s}n_{s}g_{is}\,,
\label{1}
\end{equation}
where $e_{s}$ is the charge on an ion $s$ at {\bf r}$_{2}$, $n_{s}$ is the mean number
density of  ions of type $s$, $\varepsilon_\text{r}$ is the solvent relative permittivity and
$\varepsilon_{0}$ is the vacuum permittivity. One technique to express the pair distribution
function in terms of mean electrostatic potentials is to use Kirkwood's charging process
\cite{kirkwood1}. Taking the ion $j$ to have a charge $\lambda _{2}e_{j}$  ($0 \leqslant \lambda _{2}\leqslant  1$),
then the charging process gives \cite {outh1,outh2}
\begin{equation}
g_{ij} = g_{ij} (\lambda_{2} = 0)\exp\left\{-\beta e_{j}\left[\psi (1;2) + \int_{0}^{1}\phi (1,2;2)\rd\lambda _{2}\right]\right\},  
\label{2}                           
\end{equation}
where $\phi (1,2;3)$ is the fluctuation potential defined by
\begin{equation}
\psi (1,2;3) = \psi (1;3) + \psi (2;3) + \phi (1,2;3),
\end{equation}
with $\psi (1,2;3)$ the mean electrostatic potential at {\bf r}$_{3}$ for ions $i$ and $j$
fixed at {\bf r}$_{1}$ and {\bf r}$_{2}$, respectively.  Thus, the fluctuation potential
$\phi (1,2;3)$ describes the departure from linear superposition of the singlet potentials.
Equation~(\ref{2}) for $g_{ij}$ is not symmetric, but a symmetric pair distribution function is
obtained by charging up the ion $i$ and combining the result with equation~(\ref{2}) \cite{outh3}.

    The MPB theory rests upon the closure introduced in \cite{outh4}.  Consider the
conditional potential of mean force $W(1,2;3)$, which measures the work done in bringing an
ion $s$ to {\bf r}$_{3}$, given the fixed ions $i$ and $j$ at {\bf r}$_{1}$ and {\bf r}$_{2}$,
respectively. In an analogous fashion to that for the mean electrostatic potentials, it can be
written as follows:
\begin{equation}
W(1,2;3) = W(1,3) + W(2,3) + w(1,2,3),
\end{equation}
where $w(1,2,3)$ is the corresponding departure from linear superposition of the singlet potentials
of mean force. The MPB closure is
\begin{equation}
w(1,2,3)= e_{s}\phi (1,2;3),
\end{equation}
which is analogous, but at the next hierarchical level, to the DH closure
$W(1;3) = e_{s}\psi (1;3)$. Attempts to improve the DH theory by using the closure
$W(1,2;3) = e_{s}\psi (1,2;3)$ fails as this closure neglects terms of the order being
calculated \cite{outh1,outh2,outh4,kirkwood2}.

	The governing set of differential equations for the fluctuation potential can be
derived by using the Poisson equation for one or two fixed ions. This gives for the PM
a two sphere potential problem which enables an approximate solution to be found for
$\phi (1,2;3)$. Given an appropriate fluctuation potential, an improved pair distribution
function can now be found.  Previous analysis \cite{outh1,outh2,outh3} has derived, for
a RPM,
\begin{equation}
g_{ij} = g_{ij}^{0} \exp\left\{-\frac{\beta}{2}[e_{j} L(u_{i})+e_{i} L(u_{j} )]\right\},                                                          \end{equation}
where $g_{ij}^{0} = g_{ij} (e_{i} = e_{j} = 0)$,
\begin{equation}
L(u)= \frac{1}{2r(1+y)}\left[u(r+a)+u(r-a)+ \kappa \int_{r-a}^{r+a}u(R)\rd R\right].
\end{equation}
Here, $u = r \psi(1;2)$, $r = r_{ij}$, $y = \kappa a$ with
$\kappa = [\beta/(\varepsilon_{0}\varepsilon_\text{r})\sum_{s} e_{s}^{2} n_{s}]^{1/2}$
the DH parameter. Substituting for $g_{ij}$ in equation~(\ref{1}) gives the RPM
non-linear MPB equation.

\subsection{The linear MPB and MSA equations}

    The linear form of the MPB equation, for the special case of a single
electrolyte with equal valences, is as follows:
\begin{equation}
\frac{\rd^{2}u(r)}{\rd r^{2}} = g_{ij}^{0}\kappa ^{2} r L(u),  \quad  r \geqslant a, 
\label{8}
\end{equation}
with
\begin{equation}
u(r) = \frac{r}{a}u(a)-\frac{e_{i} (r-a)}{4\piup \varepsilon_{0}\varepsilon_\text{r} a}\,, \quad 0 \leqslant r \leqslant a.
\label{9}
\end{equation}
Substituting the linear solution equation~(\ref{9}) into equation~(\ref{8}) means that
\begin{align}
L(u) & =  \frac{\kappa ^{2}}{4(1+y)} \left[2u(r+a)+(r-2a)(2+2y-\kappa r) \left(\frac{\rd u}{\rd r}\right)_{r = a}
+2(1+2y-\kappa r)u(a) + 2\kappa \int_{a}^{r+a}u(R)\rd R\right], \nonumber\\ 
& \hspace{10cm}  a \leqslant r \leqslant 2a.  
\label{10}
\end{align}
Taking the Laplace transform of equation~(\ref{8}), with $g_{ij}^{0} = 1$, gives the
general solution to be of the form
\begin{equation}
u(r) = \sum_{n=1}^{\infty} A_{n} \exp \left(-\frac{z_{n} r}{a}\right),
\label{11}
\end{equation}
where the sum is over the roots $z_{n} = \alpha _{n} + \ri\beta _{n}$
of the transcendental equation
\begin{equation}
z \cosh(z) + y \sinh(z)= z^3 \frac{(1+y)}{y^{2}}.
\label{12}
\end{equation}

The roots with the smallest real part correspond to the physical situation.
For small $y$, there are only two real roots with the smallest corresponding
to the DH solution. As $y$ increases, the two real roots coalesce at
$y_\text{c} = 1.2412$. Above this value, the two roots move off as a complex conjugate
pair until they become purely imaginary at $y_\text{I} = 7.83$.

\begin{figure}[!b]
\centerline{\includegraphics [width=0.62\textwidth]{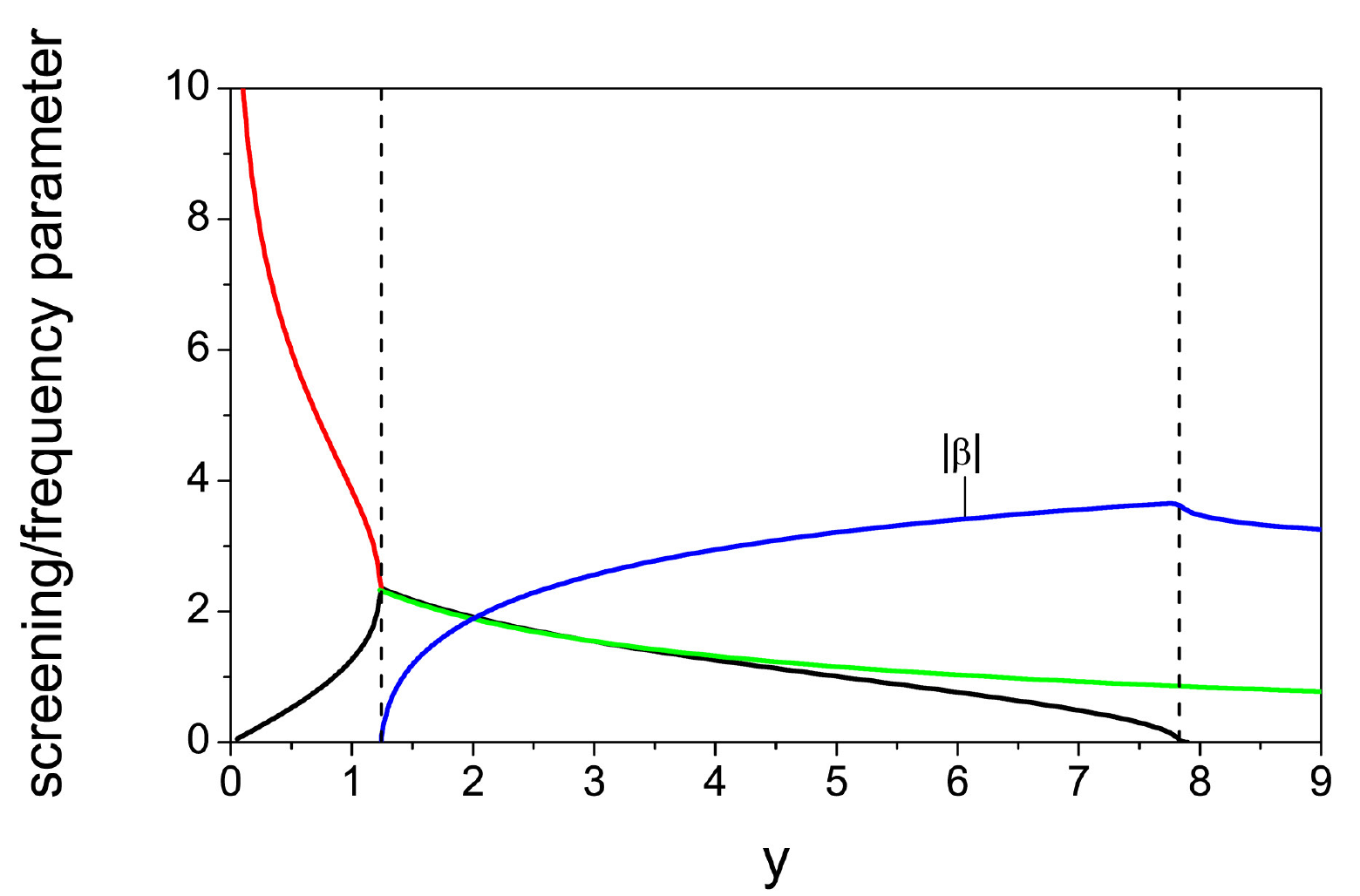}}
\caption{(Colour online) Roots of the MPB transcendental equation~(\ref{12}). For $y < y_\text{c}$
there are two real roots $\alpha _{1}$, $\alpha _{2}$ with the black (lower) curve corresponding
to the DH $\kappa$ $(= \alpha _{1}/a)$ for low $y$. For $y_\text{c} < y < y_\text{I}$, the black curve
represents the screening parameter $\alpha $, and the blue curve represents the screening frequency
$|\beta |$. The MSA roots of equation~(\ref{15}) are nearly identical to those of the MPB for
$y < y_\text{c}$, where now $y_\text{c} = 1.229$, and are hence not shown. For $y > y_\text{c}$, the MSA
screening parameter $\alpha $ is given by the upper green curve. The MSA does not have a finite
$y_\text{I}$. The two vertical dashed lines at $y_\text{c} = 1.2412$ and $y_\text{I} = 7.831$ indicate the
two MPB critical values.}
\label{fig1}
\end{figure}

	The MSA theory is one of the most successful and widely used theories of charged
fluid systems. Although of a linear nature, its appeal lies in its analytical results,
which have been applied to numerous electrolyte models and theories. The mean
electrostatic potential has been studied in \cite{outh5}, with
\begin{equation}
\frac{\rd^{2}u(r)}{\rd r^{2}} = \kappa ^{2} r L_\text{M}(u), \quad  r \geqslant a,                                                                                    
\end{equation}
where now
\begin{equation}
L_\text{M} = \frac{1}{2r}\left\{(1-b)^{2} [u(r+a)+u(r-a)] + \frac{2b}{a}\int _{r-a}^{r+a}\left(1- \frac{b}{a} |r-R|\right)u(R)\rd R\right\},
\end{equation}
with $b = [y+1-\sqrt{(1+2y)}]/y$.  The general solution is the form of
equation~(\ref{11}), where now $z$ satisfies the transcendental equation
\begin{equation}
z^{2} - 2a\Gamma z + 2(a\Gamma )^{2} [1-\exp{(z)}] = 0,
\label{15}
\end{equation}
with $a\Gamma = y(1-b)/2$. There are two real roots for small $y$ with
the smallest corresponding to the DH solution, analogous to the MPB theory. Again,
as $y$ increases, the two real roots coalesce, but at the slightly smaller value of
$y_\text{c} = 1.229$ \cite{outh5,hafskjold}. Above this critical value of $y$, the two roots
become a complex conjugate pair, but unlike the MPB situation, they do not become
purely imaginary for large $y$ \cite{blum}. Figure~\ref{fig1} displays the behaviour of the MPB and MSA
roots, while a table of some of the roots is given in \cite{outh1}. A critical value of
$y_\text{c} = 1.032$ was first predicted by Kirkwood \cite{kirkwood1} in his analysis of the
potentials of mean force. His transcendental equation, corresponding to those of
equations~(\ref{12}) and (\ref{15}), has a behaviour pattern similar to the MPB theory. Unfortunately,
his value of $y_\text{I} = 2.79$ means that his theory is restricted to much lower concentrations than
the MPB and MSA theories. Critical values of $y_\text{c}$ have been predicted by other
theories \cite{martynov1,martynov2,rasaiah,kjellander,lee2}.

We now consider $u(r)$ to be given by the two roots with the smallest real parts so that
\begin{equation}
u(r) = A_{1} \exp{\left(-\frac{z_{1} r}{a}\right)} + A_{2} \exp{\left(-\frac{z_{2} r}{a}\right)}, \quad r \geqslant 2a,
\label{16}
\end{equation}
with the first term corresponding to the DH value for small $y$. Both the MPB and MSA
predict, above their respective $y_\text{c}$, that the solution is damped oscillatory \cite{outh1,outh2}.
So, taking the complex conjugate pair $z_{1}$, $z_{2}$ to be $\alpha \pm \ri\beta$, equation~(\ref{16})
can then be written as follows:
\begin{equation}
u(r) = A \exp{\left(-\frac{\alpha r}{a}\right)}\cos{\left(\frac{\beta r}{a}-B\right)},  \quad  r \geqslant 2a,
\label{17}
\end{equation}
where $A_{1} = \bar{A}_{2} = X + \ri Y$,   $A = 2 \sqrt{X^{2} + Y^{2}}$,
$\tan{B} = Y/X$. Equation~(\ref{17}) holds for the MPB when $1.2412 <y< 7.83$,
while for the MSA when $y> 1.229$. Analysis of the Ornstein-Zernike equation indicates
that the asymptotic form of the correlation function takes on a similar form \cite{leote}.
Equation~(\ref{17}) predicts, for both theories, that the screening $\alpha /a$ decreases and
the frequency parameter $\beta /a$ increases for increasing $y$.

The approximate solution~(\ref{16}) involves two unknown constants. These cannot be
determined in a unique fashion as equation~(\ref{16}) is simply the two leading terms of
the general solution~(\ref{11}). A consistent analysis requires that the general solution
satisfies the integro-difference differential equation in $a \leqslant r \leqslant 2a$ and the
continuity of $u(a)$ and $(\rd u/\rd r)_{r = a}$. In this general case, the boundary
conditions at $r = a$ give the neutrality condition
\begin{equation}
\sum _{n=1}^{\infty}A_{n} (1 + z_{n}) \exp{(-z_{n})} = \frac{e_{i}}{4\piup \varepsilon _{0}\varepsilon_\text{r}}.                                                      
\end{equation}

Another general condition is that of Stillinger-Lovett \cite{stillinger}, which is
true away from the critical region of the RPM electrolyte. In terms of the mean electrostatic
potential, the condition is \cite{outh6}
\begin{equation}
\beta \sum_{s}e_{s} n_{s}\int \psi _{s}\rd V = 1.                                                                                      
\end{equation}

    Assuming that the solution~(\ref{16}) holds for $r \geqslant a$, then neutrality and the
Stillinger-Lovett condition allows $A_{1}$ and $A_{2}$, or equivalently $A$ and $B$,
to be calculated. Another approach, following Kirkwood \cite{kirkwood1}, is for the
solution to satisfy neutrality and equation~(\ref{16}) at $r = a$. A third possibility,
amongst others, is to evaluate $u(r)$ in $a \leqslant r \leqslant 2a$
 by substituting the solution~(\ref{16})
for $r \geqslant 2a$ into equation~(\ref{8}) and integrating twice. In this case, the boundary
conditions are $u(r)$, $\rd u(r)/\rd r$ continuous at $r = a$ and $2a$, with neutrality and
the Stillinger-Lovett condition satisfied. The three approximations are called LMPB1,
LMPB2, and LMPB3, respectively. See appendix~\ref{app} for details. The above approximate analysis
to determine the LMPB mean electrostatic potential is unnecessary for the MSA as an
alternative approach gives an analytical result \cite{outh5,henderson}. However, this
alternate formulation does not give immediately a direct identification of the MSA screening
and frequency parameters.

\section{Results and discussion}

\begin{figure}[!t]
\centerline{\includegraphics[width=0.62\textwidth]{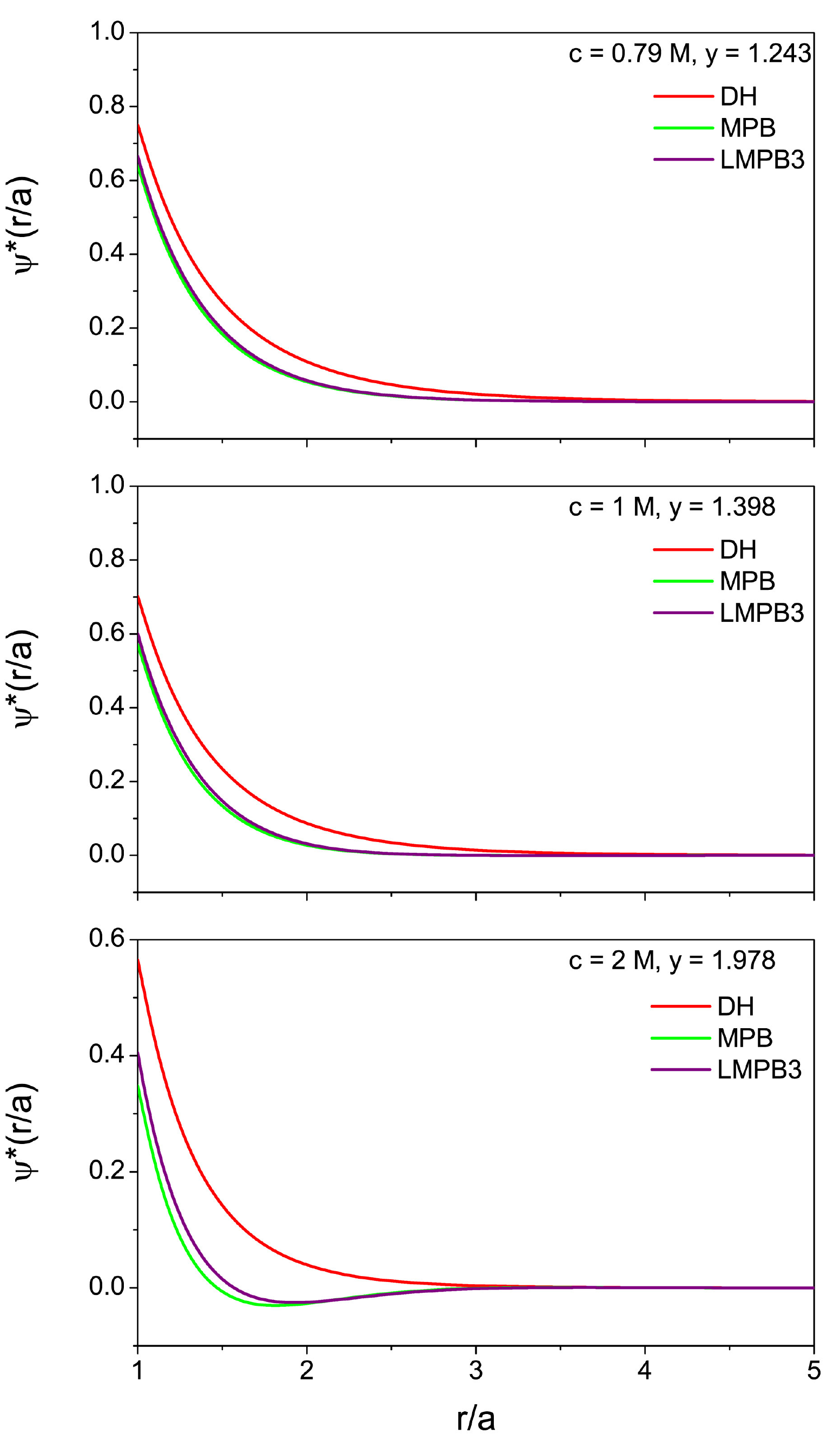}}
\caption{(Colour online) The DH, MPB and LMPB3 reduced mean electrostatic
potential $\psi ^{*}(r/a)$ $[=\beta |e|\psi (r/a)]$ as a function of $r/a$ for a 1:1 RPM
electrolyte at $y = 1.243$ ($c = 0.79$~mol/dm$^{3}$) (top panel), $y = 1.398$
($c = 1$~mol/dm$^{3}$) (middle panel), and $y = 1.987$ ($c = 2$~mol/dm$^{3}$) (bottom panel).
The LMPB1, LMPB2 and MSA results are very close to the LMPB3 result, and are not shown.}
\label{fig2}
\end{figure}

\begin{figure}[!t]
\centerline{\includegraphics[width=0.62\textwidth]{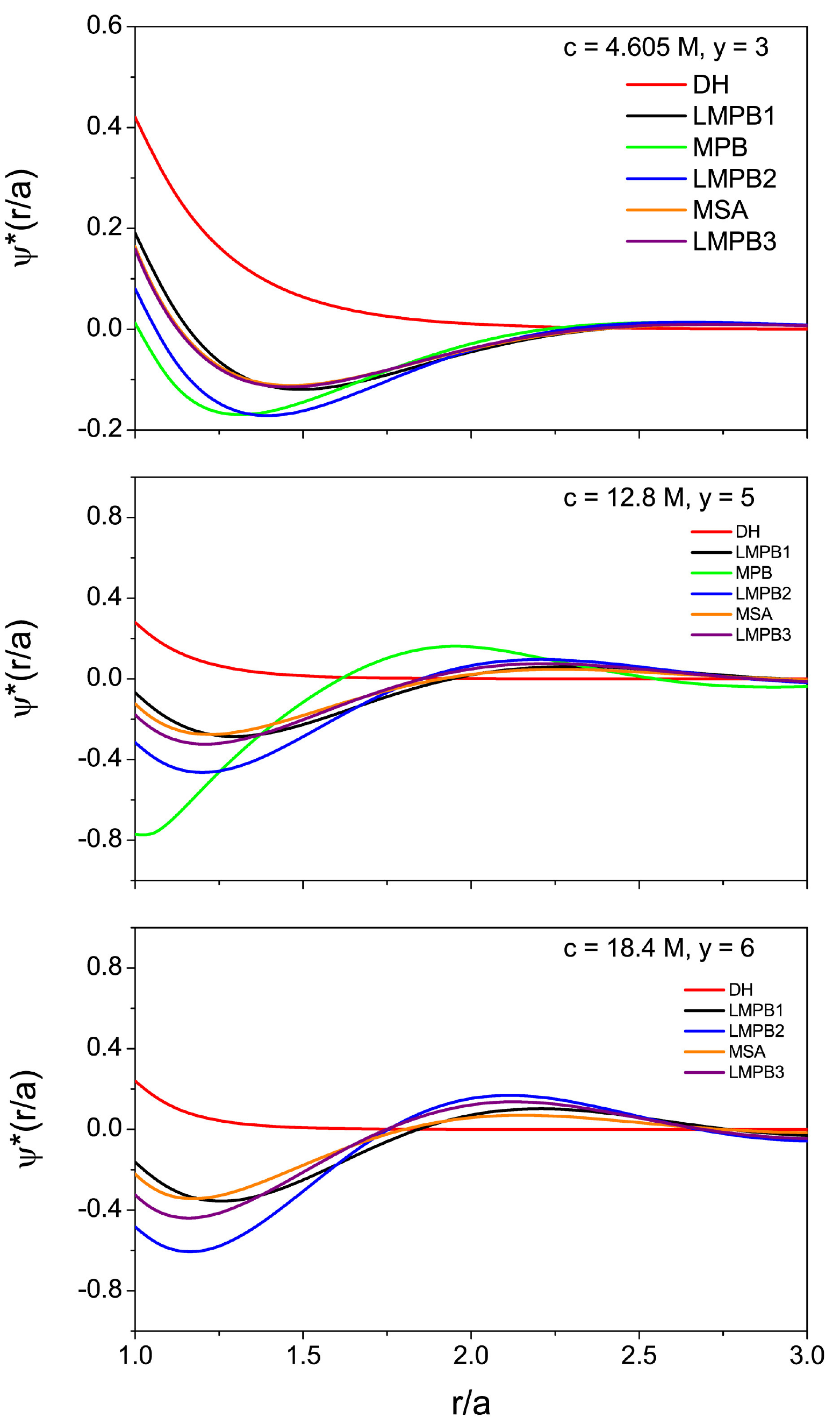}}
\caption{(Colour online) The DH, MPB, LMPB1, LMPB2, LMPB3 and MSA reduced mean
electrostatic potential $\psi ^{*}(r/a)$ $[=\beta |e|\psi (r/a)]$ as a function of $r/a$ for a 1:1
RPM electrolyte at $y = 3$ ($c = 4.605$~mol/dm$^{3})$ (top panel), $y = 5$ ($c = 12.8$~mol/dm$^{3}$)
(middle panel), and $y = 6$ ($c = 18.4$~mol/dm$^{3}$) (bottom panel). In the order from the
lowest $\psi ^{*}$(1) value, the theories are MPB, LMPB2, LMPB3, MSA, LMPB1 and DH.
In the top panel, the MSA and LMPB3 overlap and in the bottom panel there is no MPB numerical solution.}
\label{fig3}
\end{figure}

The physical parameters used in all the calculations were as follows:
the temperature $T = 298.15$~K, $a = 4.25\times 10^{-10}$~m, relative
permittivity $\varepsilon _\text{r} = 78.381$ with the concentration varying from
$c = 0.79$~M to 18.4~M, so the range of $y$ is 1.243 to 6. Plots of the
dimensionless mean electrostatic potential $\psi ^{*}(r/a)$ $[=\beta |e|\psi (r/a)]$
for the 6 theories MPB, LMPB1,
LMPB2, LMPB3, MSA and DH are shown in figures~\ref{fig2} and \ref{fig3} for a uni-univalent
electrolyte. Figure~\ref{fig2} illustrates the behaviour of $\psi ^{*}$ as $y$ increases
through  $y_\text{c}$ from $y = 1.243$ to $y = 1.978$, and figure~\ref{fig3} shows the behaviour
at the  higher values of $y = 3, 4, 6$. The non-linear MPB has been shown
to  accurately predict  both the structural and thermodynamic properties of
the RPM for 1:1  electrolytes, and thus is a good metric to assess the accuracy
of the linear  theories. For $y < y_\text{c}$, except for $y$ approaching $y_\text{c}$,
all the theories  are qualitatively very similar to the DH result, and hence,
are not shown here.  Above $y_\text{c}$, the theories demonstrate a damped
oscillatory behaviour which  cannot be predicted by the DH theory. Overall
LMPB2 is in closest agreement  with MPB, while surprisingly LMPB3 does
relatively poorly. It would have been  expected that the LMPB3 would be the
most accurate because of the treatment of the region $a \leqslant r \leqslant 2a$
via equation~(\ref{10}). Comparison with the MPB, when the exclusion volume term
is unity,  brings  the  LMPB3  and  MPB into  almost  quantitative  agreement
for    $c \leqslant  18.4$~M. This indicates that the improved mean electrostatic
potential treatment of the LMPB3  analysis needs to be balanced by incorporation of the uncharged hard  sphere term  in equation~(\ref{8}). An
interesting feature of the LMPB3 is that  besides damped oscillation terms,
there is a $(r-2a)$ quadratic contribution in the region $a \leqslant r \leqslant 2a$.

\begin{figure}[!t]
\centerline{\includegraphics[width=0.62\textwidth]{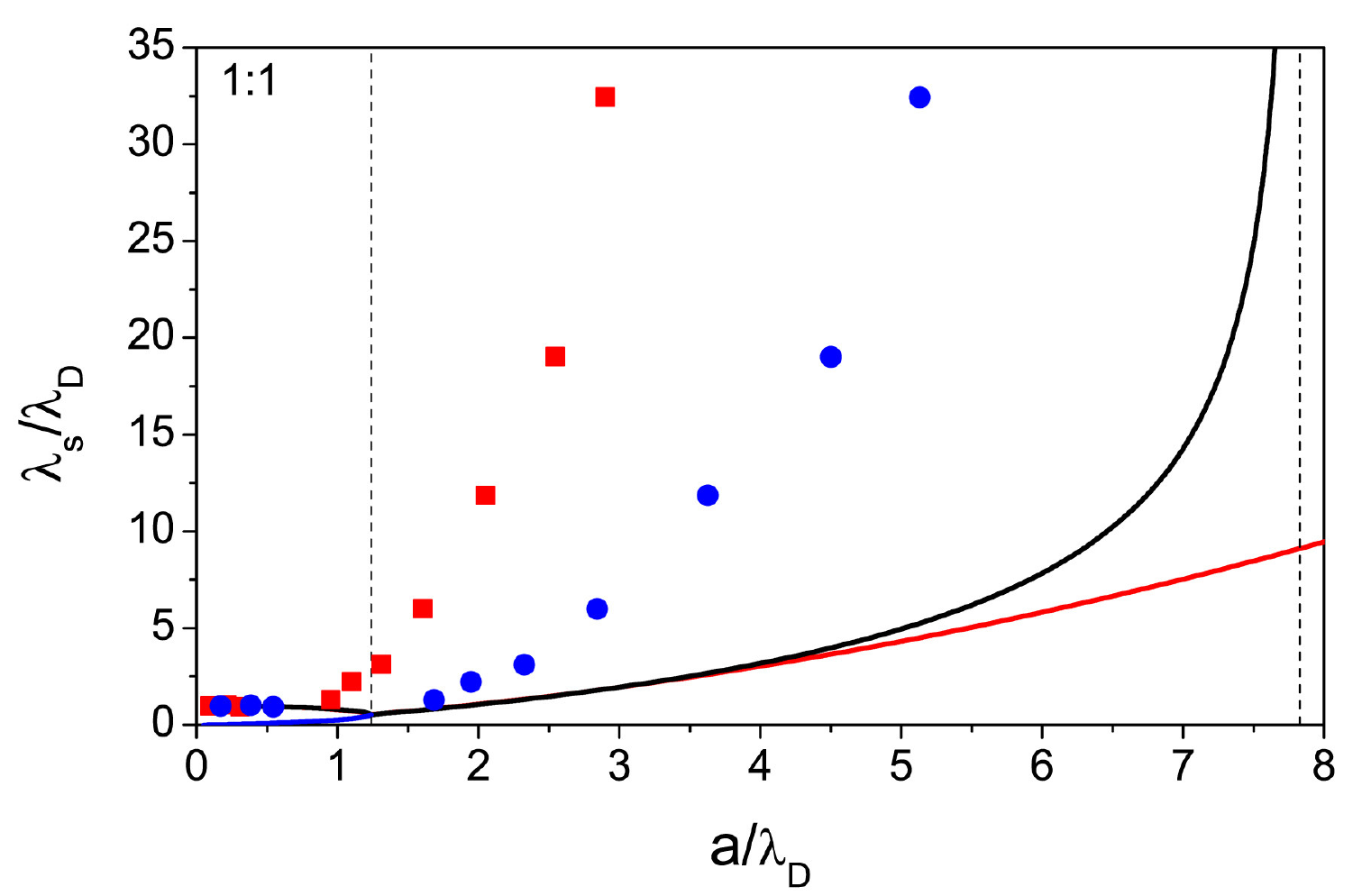}}
\caption{(Colour online) The experimental ratio $\lambda _{s}/\lambda _\text{D}$
for an aqueous solution of NaCl as functions of $a/\lambda _\text{D}$ compared with the MPB
$y/\alpha _{1}$, $y/\alpha _{2}$ for $y < y_\text{c}$, $y/\alpha $ for $y_\text{c} \leqslant  y  \leqslant  y_\text{I}$.
The filled squares and the filled circles are the experimental results for the
unhydrated ($a = 2.94 \times  10^{-10}$~m) and hydrated ($a = 5.2 \times  10^{-10}$~m) NaCl,
respectively. The experimental results are taken from the Supporting Information of Smith~A.M., Lee~A.A., Perkin~S.; electronic link given in reference \cite{lee}. The upper curve for
$y < y_\text{c}$ is that for $\alpha _{1}$ which would be unity if it was the DH value. The MSA screening
length is given by the lower red curve and is indistinguishable from the MPB value on the graphical
scale for $y  \leqslant   4$. The MSA does not have a finite $y_\text{I}$. The two vertical dashed lines at
$y_\text{c} = 1.2412$ and $y_\text{I} = 7.831$ indicate the two MPB critical values.}
\label{fig4}
\end{figure}

Experimental results for a wide class of salts indicate that the decay
length $\lambda _{s}$ of concentrated electrolytes increases with concentration,
in contrast to that of the DH prediction. The LMPB and MSA also predict that
their decay length $a/\alpha $, which corresponds to the experimental $\lambda _{s}$,
increases at higher concentrations, but at too low a rate --- see figure~\ref{fig4}.
Lee et al. \cite{lee} have analysed the
electrolyte screening length behaviour at high concentration using a scaling analysis.
They find that the decay length increases linearly with the Bjerrum length, or
alternatively  $\lambda _{s}/\lambda _\text{D} \sim (a/\lambda _\text{D})^{3}$ where
$\lambda _\text{D} = 1/\kappa $, in general  agreement with experimental results
for a range of 1:1 valency ionic liquids. We note that
in the present paper $\lambda _{s}/\lambda _\text{D} = y/\alpha $
and $a/\lambda _\text{D} =y$ so that the experimental trend thus provides a
test of theories through a study of $y/\alpha \sim  y^{3}$. Unfortunately, no
realistic comparisons can be made with the MPB and MSA predictions due
to the decay length being too small. For example, at $y = 3$, the LMPB
$y/\alpha = 1.95$, which is an order of magnitude too small (see figure~\ref{fig1}
of \cite{lee}). Only for $y$ close to $y_\text{I}$ is the LMPB decay length of
the correct order of magnitude. The most
probable reason for the shortfall is the lack of any treatment of the solvent.
At low to medium solution concentration, the solvent contribution can be partially
treated by using an hydrated ion diameter. In figure~\ref{fig4}, the experimental ratio
$\lambda _{s}/\lambda _\text{D}$ for varying $a/\lambda _\text{D}$ is given for an aqueous
solution of NaCl for $a = 2.94\times 10^{-10}$~m and $a = 5.2 \times  10^{-10}$~m.
The use of the hydrated diameter $a = 5.2 \times  10^{-10}$~m brings the
experimental points closer to the MPB results. The high concentration of the
solvent relative to the solute ions means that the structure around the ions
is driven by the solvent. A step in this direction is to consider the solvent
primitive model where the solvent is modelled by uncharged hard spheres moving
in a constant permittivity. Recently, the underscreening in this solvent primitive
model has been studied in the MSA \cite{rotenberg}. The ratio $\lambda_{s}/\lambda_\text{D}$
increases at higher concentrations as required by experiment, but at too slow a rate.

\section{Summary}

The potential approach of the MPB theory provides a natural extension to the DH theory.
At lower electrolyte concentrations, the predictions mimic the DH theory but qualitative
differences occur at higher concentrations. In the region $y_\text{c} < y < y_\text{I}$, the MPB predicts a
damped oscillatory potential which is in accordance with simulation and other theoretical work.
For this interval of $y$, the linear potential is of the form
$u(r) = A \exp{(-\alpha r/a)}\cos{(\beta r/a-B)}$,    $r \geqslant  2a$, where $A$, $B$ are constants
and $\alpha $, $\beta $ given by the solution of equation~(\ref{12}). There is no unique way of determining the
constants, so three methods are considered, in each case the neutrality condition being
satisfied. The Stillinger-Lovett condition is used in both the LMPB1 and LMPB3 approaches
with the solution holding at $r = a$ and $r = 2a$, respectively, while the LMPB2 is derived by
assuming that the equation is satisfied at $r = a$. Comparison of the three theories with
the accurate non-linear MPB for 1:1 salts indicates that overall the simple LMPB2 provides
the best approximation. The relative poor behaviour of LMPB3 is deceptive. The LMPB3
potential is nearly identical to that of the non-linear MPB when its exclusion volume
term is unity. This means that realistic improvements to the linear MPB lie through
treating the exclusion volume term in the LMPB3 approach.

Experimental results have shown that the screening parameter, as interpreted via a
DH type potential, initially increases as the electrolyte concentration rises, but then
decreases at higher concentrations. In the LMPB theory at lower concentrations,
there are two pure damped exponential terms with screening parameters $\alpha _{1}/a$, $\alpha _{2}/a$ with
the smaller $\alpha _{1}/a$ corresponding to the DH screening. The parameters $\alpha _{1}$, $\alpha _{2}$ coalesce at
$y_\text{c}$ as the concentration increases and become complex conjugate roots $\alpha \pm \ri\beta $ for
$y_\text{c} < y < y_\text{I}$. These complex conjugate roots lead to a damped oscillatory behaviour
for the electrostatic potential with screening parameter $\alpha /a$. In this region of $y$,
the screening parameter reduces with concentration, as seen in experiments over a
similar range of $y$. The experimental results also suggest that the decay length
increases linearly with the Bjerrum length. Such a behaviour is not seen in the
LMPB as the rate of a decrease of the screening length is too small, only at values
of $y$ close to $y_\text{I}$ is the screening length of the correct order of magnitude. The
MSA mean electrostatic potential has features in common with the LMPB. It becomes
damped oscillatory at a value of $y_\text{c}$ slightly less than that of the MPB, but in
contrast there is no finite value of $y_\text{I}$. This means that its screening factor
behaves in an analogous fashion to that of the MPB as $y$ increases through $y_\text{c}$.
Above $y_\text{c}$, the MSA screening parameter is greater than that of the MPB and slowly reduces
to zero as $y$ tends to infinity. Close agreement with experiment at the higher electrolyte
concentrations is unlikely with the present PM model. A critical feature of the
PM is the lack of any adequate treatment of the solvent. Models such as the
solvent primitive model have provided insight as to the importance of solvent
steric effects, but both the solvent and solute molecules need to be adequately
modelled and treated on an equal footing.

The linear MPB equations studied here complement the full,
non-linear MPB approach to the electrolyte solution theory. From a theoretical
perspective, these solutions offer a valuable physical insight by painting a
clearer picture of the onset of oscillations in the mean electrostatic profile
and screening in electrolytes.  The LMPB solutions transcend the simplistic DH
view, compare favourably with the more well known MSA while retaining the
simplicity and convenience of their analytic nature. Thus, the LMPB
theories might well prove useful as a basis for interpreting the thermodynamic
properties of concentrated electrolytes. Again, the theories could be applied
to studying the liquid-gas transition in simple electrolytes in an analogous
fashion to the MSA. Although the LMPB1 and LMPB3 satisfy the Stillinger-Lovett
condition, this does not prohibit the LMPB1 and LMPB3 from having a solution
in the liquid-gas region, as seen with the MSA. The solution of the LMPB theories
is basically dependent on $y$ through the soluton of equation~(\ref{12}), so the theories
can be solved in the simulated liquid-gas coexistence region. The pattern of the LMPB
results for the range of concentration treated suggests these to be
potentially very useful as initial input in the numerical solution of the
full MPB equation and alternative non-linear theories.

\appendix
\section{Analytic expressions for the three LMPB theories}
\label{app}

\subsection{$0  \leqslant y \leqslant 1.2412$}
\label{A1}

Roots of equation (\ref{12}), $z_{1} = \alpha _{1}$, $z_{2} = \alpha _{2}$.
\begin{align*}
\psi _{i} = \left[A_{1}\exp(-\alpha _{1}r/a)+A_{2}\exp(-\alpha _{2}r/a)\right]/r,  \hspace{0.3cm}   r \geqslant a.
\end{align*}

{\bf (a)} LMPB1
\begin{align*}
A_{1}  &=  \frac{e_{i}}{4\piup \varepsilon _{0}\varepsilon _\text{r}}\exp(\alpha _{1})[\alpha _{1}^{2}G_{2}-\omega (1+\alpha _{2})]/D,    \\
A_{2}  &= \frac{e_{i}}{4\piup \varepsilon _{0}\varepsilon _\text{r}}\exp(\alpha _{2})[-\alpha _{2}^{2}G_{1}+\omega (1+\alpha _{1})]/D,     \\
D  &=  \alpha _{1}^{2}(1+\alpha _{1})G_{2}-\alpha _{2}^{2}(1+\alpha _{2})G_{1}\,, \hspace{0.5cm} \omega = 6\alpha _{2}^{2}/y^{2},  \\
G_{j}  &=  \alpha _{j}^{3}+3\alpha _{j}^{2}+6\alpha _{j}+6, \hspace{0.5cm}  j = 1,2.     
\end{align*}

{\bf (b)} LMPB2, see also reference \cite{outh7}.
\begin{align*}
A_{1}  &=  \frac{e_{i}}{4\piup \varepsilon _{0}\varepsilon _\text{r}}b_{2}\exp(\alpha _{1})/D,   \\
A_{2}  &=  -\frac{e_{i}}{4\piup \varepsilon _{0}\varepsilon _\text{r}}b_{1}\exp(\alpha _{2})/D ,  \\
D  &=  b_{2}(1+\alpha _{1}) - b_{1}(1+\alpha _{2}),   \\
b_{j} & =  \frac{2\alpha _{j}^{2}}{\lambda }-(2+y)\alpha _{j}-2(1+y)-\frac{2y}{\alpha_{j}}-2\left(1-\frac{y}{\alpha _{j}}\right)\exp(-\alpha _{j}),  \hspace{0.5cm} j = 1,2,               \\
\lambda  &=  \frac{y^{2}}{2(1+y)}.             
\end{align*}

{\bf (c)} LMPB3. Not given as results very detailed, see for example subsection~\ref{A2} below.

\subsection{$1.2412 < y < 7.83$}
\label{A2}

Roots of equation (\ref{12}), $z_{1} = \alpha + \ri\beta$,  $z_{2} = \alpha -\ri\beta$.
\begin{align*}
\psi _{i} &= (e_{i}/4\piup \varepsilon _{0}\varepsilon _\text{r}r)A\exp[-\alpha (r/a-1)]\cos[\beta(r/a-1)-B], \quad r \geqslant a,\\
A &= \sqrt{X^{2}+Y^{2}}/D,  \quad  B = \tan ^{-1}(Y/X),\\
R &= \alpha ^{2} - \beta ^{2}, \quad   S = \alpha ^{2} + \beta ^{2}.
\end{align*}

{\bf (a)} LMPB1
\begin{align*}
G  &= G_{1} + \ri G_{2}\,,   \\
G_{1} & = \alpha ^{3} -3\alpha \beta ^{2}+3\alpha ^{2}-3\beta ^{2} +6\alpha +6,  \\
G_{2} & = 3\alpha ^{2}\beta -\beta ^{3} +6\alpha \beta +6\beta,     \\
H & = \frac{6}{y^{2}}S^{2},  \\
X & = \beta H-RG_{2}+2\alpha \beta G_{1}\,,
\end{align*}
\begin{align*}
Y & = (1+\alpha )H-RG_{1}+2\alpha \beta G_{2}\,,    \\
D & = (1+\alpha )X-Y\beta.    
\end{align*}

{\bf (b)} LMPB2

From \ref{A1}~(b), $b_{1}=Y+\ri X$,  $b_{2}=Y-\ri X$, giving
\begin{align*}
Y & = \frac{2R}{\lambda}-(2+y)\alpha -2(1+y)-\frac{2y\alpha }{S}-2\exp(-\alpha )(p\cos\beta +q\sin\beta ),   \\
X & = \frac{4\alpha \beta }{\lambda }-(2+y)\beta +2q-2\exp(-\alpha )(q\cos\beta  -p\sin\beta  ),    \\
D & = (1+\alpha )X-Y\beta ,    \\
\lambda  & = \frac{y^{2}}{2(1+y)}\,, \hspace{0.25cm} p=1-\frac{y\alpha }{S}\,,  \hspace{0.25cm} q=\frac{qy}{S}. 
\end{align*}

{\bf (c)} LMPB3
\begin{align*}
\psi _{i} & =  \frac{\lambda e_{i}}{4\piup \varepsilon _{0}\varepsilon _\text{r}}[DX+EY+\mu (r)+c_{8}\eta (r)], \hspace{0.25cm} a \leqslant r  \leqslant 2a,                   \\
\psi _{i} & =  \frac{e_{i}}{4\piup \varepsilon _{0}\varepsilon _\text{r}} A \exp(-\alpha r/a)\cos{(\beta r/a -B)}, \hspace{0.25cm} r \geqslant 2a,            \\
A & =  2\sqrt{X^{2}+Y^{2}}, \hspace{0.25cm}  B = \tan ^{-1}(Y/X),   \\
\lambda & =  \frac{y^{2}}{2(1+y)}\,,    \hspace{0.25cm}  p = 1-\frac{y\alpha }{S}\,,   \hspace{0.25cm}  q = \frac{y\beta }{S}\,,  \\
R & =  \alpha ^{2} - \beta ^{2},   \hspace{0.25cm}   S = \alpha ^{2} + \beta ^{2},     \\
x & =  \frac{r-2a}{a}\,,     \\         
c_{1} & = \frac{6y^{3}}{12+12y-y^{3}}\,,              \nonumber   \\
c_{2} & = \frac{120y^{2}}{240+240y+y^{2}(60+15c_{1}+40y+4yc_{1})}\,,              \nonumber \\
c_{3} & = \frac{5c_{1}}{24}\,,   \nonumber  \\
c_{4} & = \frac{1}{24}(4c_{3}+5)+\frac{y}{120}(5c_{3}+6)+d_{2}c_{8}\,,  \nonumber    \\
c_{5} & = 1-\frac{y^{2}(3+2c_{8})}{6}-\frac{y^{4}c_{4}}{2(1+y)}\,,    \nonumber    \\
c_{6} & = \frac{19}{120}+\frac{29y}{720}+\frac{c_{3}}{8}+\frac{yc_{3}}{30}\,,   \nonumber   \\
c_{7} & = 1-y^{2}\left[\frac{3}{2}+\frac{2y}{3}+\frac{c_{1}(4+y)}{16}+c_{8}c_{9}\right],   \nonumber   \\
c_{8} & = -c_{2}\frac{24+4c_{1}+12y+yc_{1}}{24}\,,     \nonumber   \\
c_{9} & = \frac{200+110y+3c_{1}(15+4y)}{240}\,,    \nonumber   \\
d_{1} & = p\left(1+\frac{\alpha }{S}\right)+\frac{q^{2}}{y}+\frac{3y}{2}\,,    \nonumber    \\
d_{2} & = \beta \frac{y(S+2\alpha )-S}{S^{2}}\,,         \nonumber      \\
D     & = K + D_{1}+ Mx +\left[\frac{c_{1}V}{2}+\frac{y\exp(-2\alpha )B_{4}}{S}\right]x^{2} + M_{3}\eta (r),   \nonumber  
\end{align*}
\begin{align*}
E     & = L+E_{1}+ Nx +\left[\frac{c_{1}W}{2}+\frac{y\exp(-2\alpha )B_{5}}{S}\right]x^{2} + N_{3}\eta (r),    \nonumber  \\
D_{1} & = \frac{2\exp\left[-\left(\frac{r}{a}+1\right)\alpha \right](Rs_{1}+2\alpha \beta s_{2})}{S^{2}}\,,      \nonumber    \\
E_{1} & = \frac{2\exp\left[-\left(\frac{r}{a}+1\right)\alpha \right](2\alpha \beta s_{1}-Rs_{2})}{S^{2}}\,,      \nonumber    \\
s_{1} & = p\cos\left[\beta\left(\frac{r}{a}+1\right)\right]+q\sin\left[\beta\left(\frac{r}{a}+1\right)\right],   \nonumber  \\   s_{2} & = q\cos\left[\beta\left(\frac{r}{a}+1\right)\right]-p\sin\left[\beta\left(\frac{r}{a}+1\right)\right],   \nonumber \\
\eta (r) & = \frac{1}{240}\left[c_{1}(15+4y)+120+40(1-y)x-10x^{2}\right]x^{2},   \nonumber   \\
\mu (r) & = \frac{1}{48}[c_{1}(4+y)+24]x^{2}-\frac{y}{6}x^{3},   \nonumber    \\
K & = \frac{2}{\lambda }\exp(-2\alpha )\cos2\beta-\frac{2}{s^{2}}(RB_{2}+2\alpha \beta B_{3})\exp(-3\alpha ), \nonumber  \\     L & = \frac{2}{\lambda }\exp(-2\alpha )\sin2\beta-\frac{2}{s^{2}}(2\alpha \beta B_{2}-RB_{3})\exp(-3\alpha ), \nonumber  \\
B_{2} & = p\cos3\beta +q\sin3\beta, \hspace{0.5cm} B_{22} = p\cos2\beta +q\sin2\beta,   \nonumber    \\
B_{3} & = q\cos3\beta -p\sin3\beta, \hspace{0.5cm} B_{32} = q\cos2\beta -p\sin2\beta,   \nonumber    \\
B_{4} & = \alpha \cos2\beta - \beta \sin2\beta ,   \nonumber    \\
B_{5} & = \beta \cos2\beta + \alpha \sin2\beta ,   \nonumber    \\
B_{6} & = \left(1+\frac{\alpha }{S}\right)\cos2\beta -\frac{\beta }{S}\sin2\beta ,   \\
B_{7} & = \left(1+\frac{\alpha }{S}\right)\sin2\beta +\frac{\beta }{S}\cos2\beta ,   \nonumber    \\
B_{8} & = \left(2+\frac{\alpha }{S}\right)\cos3\beta -\frac{\beta }{S}\sin3\beta ,   \nonumber    \\
B_{9} & = \left(2+\frac{\alpha }{S}\right)\sin3\beta +\frac{\beta }{S}\cos3\beta ,   \nonumber    \\
\delta _{1} & = \alpha ^{3}-3\alpha \beta ^{2},   \hspace{0.25cm}  \delta _{2}=\beta ^{3}-3\beta \alpha ^{2},  \nonumber   \\
M & = \frac{2}{S}(\alpha B_{2}+\beta B_{3})\exp(-3\alpha )-\frac{2}{\lambda }(\alpha \cos2\beta +\beta \sin2\beta ) \exp(-2\alpha ),   \nonumber    \\
N & = \frac{2}{S}(\beta B_{2}-\alpha B_{3})\exp(-3\alpha )-\frac{2}{\lambda }(\alpha \sin2\beta -\beta \cos2\beta ) \exp(-2\alpha ),   \nonumber    \\
P & = \left\{\frac{y}{3S}B_{4}+\frac{2}{S^{3}}\left[\delta _{1}B_{22}-\delta _{2}B_{32}-(\delta _{1}B_{2}-\delta _{2}B_{3})
\exp(-\alpha)\right]\right\}\exp(-2\alpha),   \nonumber    \\
Q & = \left\{\frac{y}{3S}B_{5}+\frac{2}{S^{3}}\left[-\delta _{2}B_{22}-\delta _{1}B_{32}+(\delta _{2}B_{2}+\delta _{1}B_{3})
\exp(-\alpha)\right]\right\}\exp(-2\alpha),   \nonumber    \\
V & = K -\frac{M}{2}+P,   \nonumber    \\
W & = L -\frac{N}{2}+Q,   \nonumber    \\
V_{1} & = \frac{2}{S}\exp(-2\alpha)(d_{1}B_{4}+d_{2}B_{5}),   \nonumber    \\
W_{1} & = \frac{2}{S}\exp(-2\alpha)(d_{1}B_{5}-d_{2}B_{4}),   \nonumber    \\
M_{2} & = \frac{2}{S}\left\{\beta [-q\cos2\beta +(y+p)\sin2\beta ]-\alpha [(y+p)\cos2\beta +q\sin2\beta ] \right\}\exp(-2\alpha ),   \nonumber    
\end{align*}
\begin{align*}
N_{2} & = \frac{2}{S}\left\{\alpha [q\cos2\beta -(y+p)\sin2\beta ]-\beta [(y+p)\cos2\beta +q\sin2\beta ]\right\}
\exp(-2\alpha ),   \nonumber    \\
M_{3} & = c_{2}(M-c_{1}V+M_{2}),  \hspace{0.25cm}   N_{3} = c_{2}(N-c_{1}W+N_{2}),   \nonumber    \\
K_{2} & = \Big\{\frac{5y}{12S}B_{4}+\frac{2}{S^{3}}\left[\delta _{1}(pB_{6}+qB_{7})-\delta _{2}(qB_{6}-pB_{7})\right] \nonumber   \\
      & - \exp(-\alpha)[\delta _{1}(pB_{8}+qB_{9})+\delta _{2}(pB_{9}-qB_{8})]\Big\}\exp(-2\alpha),   \nonumber    \\
L_{2} & = \Big\{\frac{5y}{12S}B_{5}+\frac{2}{S^{3}}\left[-\delta _{2}(pB_{6}+qB_{7})+\delta _{1}(qB_{7}-pB_{6})\right]
\nonumber   \\
      & + \exp(-\alpha)[\delta _{2}(pB_{8}+qB_{9})-\delta _{1}(pB_{9}-qB_{8})]\Big\}\exp(-2\alpha),   \nonumber    \\
K_{3} & = \frac{3K}{2}-\frac{2M}{3}+\frac{5c_{3}V}{24}+K_{2}\,,   \nonumber    \\
L_{3} & = \frac{3L}{2}-\frac{2N}{3}+\frac{5c_{3}W}{24}+L_{2}\,,   \nonumber    \\
M_{4} & = \lambda \left(K-2M+\frac{3c_{1}V}{2}+V_{1}+c_{9}M_{3}\right) ,  \nonumber    \\
N_{4} & = \lambda \left(L-2N+\frac{3c_{1}W}{2}+W_{1}+c_{9}N_{3}\right),   \nonumber    \\
F & = \frac{2}{S}y^{2}\exp(-2\alpha )\left[2B_{4}+\frac{1}{s}(R\cos2\beta -2\alpha \beta \sin2\beta )\right],  \nonumber \\
G & = \frac{2}{S}y^{2}\exp(-2\alpha )\left[2B_{5}+\frac{1}{s}(R\sin2\beta +2\alpha \beta \cos2\beta )\right],  \nonumber \\
F_{1} & = \frac{y^{2}M_{3}}{3}+\lambda y^{2}(K_{3}+c_{6}M_{3})+F,   \nonumber    \\
G_{1} & = \frac{y^{2}N_{3}}{3}+\lambda y^{2}(L_{3}+c_{6}N_{3})+G,   \nonumber    \\
X     & = \frac{N_{4}c_{5}-G_{1}c_{7}}{D_{0}}\,,    \nonumber    \\
Y     & = \frac{F_{1}c_{7}-M_{4}c_{5}}{D_{0}}\,,    \nonumber    \\
D_{0} & = N_{4}F_{1}-G_{1}M_{4}.   \nonumber    
\end{align*}

\newpage

\ukrainianpart

\title{Коментарі стосовно лінійного модифікованого рівняння Пуассона-Больцмана в теорії розчинів електролітів
}

\author{К.Б. Оутвайт\refaddr{label1}, Л.Б. Бхуян\refaddr{label2}}

\addresses{
	\addr{label1}Факультет прикладної математики, Унiверситет Шеффiлда, Шеффiлд, Великобританiя
	\addr{label2}Лабораторiя теоретичної фiзики, Унiверситет Пуерто Рiко, Пуерто Рiко, США
}

\makeukrtitle

\begin{abstract}
 Запропоновано три аналітичні результати для лінійної форми
	модифікованого рівняння Пуассона-Больцмана в теорії об'ємних електролітів.
	Ці результати порівнюються з результатами середньо сферичного наближення.
	Лінійні теорії передбачають перехід середнього електростатичного
	потенціалу від загасаючої експоненційної поведінки типу Дебея-Гюккеля до загасаючої
осциляційної поведінки при збільшенні концентрації електроліту
	вище критичної величини. Радіус екранування зменшується зі збільшенням
	концентрації, коли середній електростатичний потенціал загасає осциляційно.
	Проведено порівняння  з одним набором нещодавніх експериментальних
	результатів по екрануванню для водних електролітів NaCl.

	\keywords концентровані електроліти, обмежена примітивна модель, радіус екранування, лінійна модифікована теорія Пуассона-Больцмана

\end{abstract}

\end{document}